# Geodesic Nature and Quantization of Shift Vector

Hua Wang[1,2]*, Kai Chang[1,2]*

[1]Center for Quantum Matter, Zhejiang University, Hangzhou 310058, China

[2]School of Physics, Zhejiang University, Hangzhou 310058, China

* Correspondence to: daodaohw@zju.edu.cn, kchang@zju.edu.cn

## Abstract

We present the geodesic nature and quantization of geometric shift vector in quantum systems, with the parameter space defined by the Bloch momentum, using the Wilson loop approach. Our analysis extends to include bosonic phonon drag shift vectors with non-vertical transitions. We demonstrate that the gauge invariant shift vector can be quantized as integer values, analogous to the Euler characteristic based on the Gauss–Bonnet theorem for a manifold with a smooth boundary. We reveal intricate relationships among geometric quantities such as the shift vector, Berry curvature, and quantum metric. Our findings demonstrate that the loop integral of the shift vector in the quantized interband formula contributes to the non-quantized component of the trace of conductivity in the circular photogalvanic effect. The Wilson loop method facilitates first-principles calculations, providing insights in the geometric underpinnings of these interband gauge invariant quantities and shedding light on their nonlinear optical manifestations in real materials.

**Introduction.** The exploration of the geometry of quantum states in understanding the nonlinear and nonequilibrium responses of electronic systems to both static and dynamic electromagnetic fields has garnered significant attention, particularly evident in phenomena such as the nonlinear anomalous Hall effect[1] and shift current response[2-4]. Nonlinear optical responses transcend the realm of traditional single-state geometric properties, revealing as deeper connection to the underlying Riemannian geometry[5]. Notably, noncentrosymmetric materials exhibit even-order nonlinear photocurrent responses when subjected to an external electromagnetic field. A prominent example is the anomalous shift of charge carrier wave packets in real space through a second-order process upon photon excitation. This phenomenon, known as shift current, elucidates a mechanism crucial to understanding the bulk photovoltaic effect and is characterized by the shift vector, an interband geometrical property bridging two quantum states. Moreover, the shift vector plays a pivotal role in diverse phenomena like Landau-Zener tunneling[6], twisted Schwinger effect[7], high harmonic generation[8], and gauge invariant formulation of the semiconductor Bloch equations[9].

Quantized quantities are both rare and captivating in electromagnetic DC responses. An example of quantized optical response is the second-order circular photogalvanic effect (CPGE) discovered in Weyl semimetals[10]. The shift vector in bulk photovoltaic effect for an arbitrary closed loop in the parameter space generally remains non-quantized. In a specific scenario involving interface reflection and a vanishing Berry connection, a quantized circulation of the anomalous shift emerges[11]. It has been demonstrated that 3D multi-gap topological insulators can exhibit quantized circular shift photoconductivities[12]. The quantization phenomenon is explained through the integrated torsion tensor and non-Abelian Berry connection, which give rise to Chern-Simons forms. Recently, the Euler class of two-band subspaces [13] and the interband characteristics in time-dependent quantum systems[14] have been investigated. Much like the intraband geometric quantities provide insights into the quantum Hall response under a static electric field, the interband character encodes fundamental information about the quantum state of the system during light-matter interactions. The quantization arises from interband transitions contributing to the optical response, highlighting the role of quantum geometry in emergent electromagnetic properties.

In this article, we elucidate the equivalence between the geometrical shift vector and the quantum geometric potential[15], demonstrating its pivotal role as geodesic curvature. Additionally, we introduce the Wilson representation for the quantized interband character and extend our analysis to bosonic phonon drag shift vector with non-vertical transitions. This quantization, expressed in integers, serves as a counterpart to the Euler characteristic number for manifolds with a smooth boundary. According to the Gauss-Bonnet theorem, the Euler characteristic number derives from two contributions: the surface integral of the Gaussian curvature and the loop integral of the geodesic curvature along the boundary. In particular, the shift vector plays the role of geodesic curvature, while the difference in Berry curvature between two levels mirrors the analogy to the Gaussian curvature. The utilization of the Wilson loop method allows us to grasp the geometric nature of the quantization and conduct rigorous first-principles calculations. We propose that the loop integral of the shift vector contributes to the non-quantized component of the trace of CPGE conductivity using a two-band model.

**Geometric shift vector and coordinate shift.** In materials lacking inversion symmetry, the shift current bulk photovoltaic effect arises as a second order nonlinear optical phenomenon. This effect induces photocurrents without the need for an externally applied static electric field. The local gauge invariant shift vector within the shift current can be expressed as[4,16]

$$R_{mn}^{a,b}(\boldsymbol{k}) = \mathcal{A}_m^a(\boldsymbol{k}) - \mathcal{A}_n^a(\boldsymbol{k}) - \partial_a \arg r_{mn}^b(\boldsymbol{k}), \quad (1)$$

Where $\mathcal{A}_n = i\langle n|\partial_k|n\rangle$ and $r_{mn} = i\langle m|\partial_k|n\rangle(1-\delta_{mn})$ are intraband and interband Berry connection for Bloch states $|m\rangle$ and $|n\rangle$, respectively. The geometric aspect of the sshift vector is intricately linked with the quantum metric and Berry curvature via the Christoffel symbols[17,18]. Conceptually, it can be understood as the disparity in the real-space charge center between two bands during a direct optical transition. Therefore, the modified band gap, denoted by $E_{\text{gap}} + e\boldsymbol{E} \cdot \boldsymbol{R}_{mn}$, provides a more accurate description of the instantaneous band gap in the presence of an electric field $\boldsymbol{E}(\omega)$. In the static limit $\omega \to 0$, this modification can induce nonreciprocal Landau-Zener tunneling, attributable to the geometric correction of the band gap. The shift vector can also be generalized to include dispersive effects associated with non-vertical optical transitions, a phenomenon referred to as the photon drag effect[19]. The semiclassical explanation for the origin of the shift vector and phonon drag or scattering shift vector[20] can be attributed to the anomalous coordinate shift, which corresponds to the scattering of the wave packet from the state with average momentum $\boldsymbol{k}$ into the one with $\boldsymbol{k}'$, given by[21]

$$\delta R_{mn}^a(\boldsymbol{k}, \boldsymbol{k}') = \mathcal{A}_m^a(\boldsymbol{k}') - \mathcal{A}_n^a(\boldsymbol{k}) - D_{\boldsymbol{k},\boldsymbol{k}'} \arg V_{mn}(\boldsymbol{k}, \boldsymbol{k}'), \quad (2)$$

where $D_{\boldsymbol{k},\boldsymbol{k}'} = \partial_{\boldsymbol{k}} + \partial_{\boldsymbol{k}'}$, and $V_{mn}(\boldsymbol{k}, \boldsymbol{k}')$ represents the scattering potential matrix elements. The type of scattering potential involved influences the phenomena arising from the coordinate shift. For instance, a defect potential $V$ can lead to the side jump mechanism of anomalous Hall effect [22]. On the other hand, when the potential originates from an electric field, we obtain the original shift vector formula. Moreover, a phonon-induced scattering potential gives rise to the phonon drag shift vector[23,24]

$$R_{mn,\nu}^a(\boldsymbol{k}, \boldsymbol{k}') = \mathcal{A}_m^a(\boldsymbol{k}') - \mathcal{A}_n^a(\boldsymbol{k}) - D_{\boldsymbol{k},\boldsymbol{k}'} \arg g_{mn,\nu}(\boldsymbol{k}, \boldsymbol{k}'), \quad (3)$$

where $g_{mn,\nu}(\boldsymbol{k}, \boldsymbol{k}')$ is the electron-phonon coupling matrix and $\nu$ is the phonon mode. The phonon drag shift vector plays a crucial role in the kinetic processes involved in the relaxation of photo-excited electrons, consequently resulting in non-vanishing shift currents in thermal equilibrium, which is overlooked in most recent studies. In certain scenarios, the relaxation of photo-excited electrons due to phonons and the recombination of electron-hole pairs may significantly outweigh the contribution of the shift current by an order of magnitude[25]. Hence, these factors should be taken into account when analyzing photovoltaic current.

**Geodesic nature of shift vector**. Geodesic curvature is a concept from differential geometry that measures the rate at which a curve α(s) deviates from being a straight line (geodesic) within a curved surface or manifold $\boldsymbol{S}$. It precisely measures the degree to which a vector function $\boldsymbol{V}(s)$ curves away from the tangent direction $\boldsymbol{T}(s) = \nabla\alpha$ as it parallel transports along the curve α(s). Illustrated in Fig. 1(a), the geodesic curvature is expressed as $\kappa_g = d\theta/ds$, where $\theta = \text{angle}\langle \boldsymbol{V}, \boldsymbol{T}\rangle$. This concept is essential for understanding the geometry of curved spaces and holds substantial importance across diverse fields including differential geometry, physics, and engineering. To elucidate the geodesic nature of shift vector and bosonic drag shift vector, we perform a gauge transformation on the cell-periodic part of a Bloch state with a Berry phase $|\widetilde{m}\rangle = e^{i\int \mathcal{A}_m(\bm{k})\cdot d\bm{k}}|m\rangle$. Subsequently, we derive the matrix elements

$$\langle \widetilde{n}|\partial_k|\widetilde{m}\rangle = e^{i\int(\mathcal{A}_m - \mathcal{A}_n)\cdot d\bm{k}}\langle n|\partial_k|m\rangle. \tag{4}$$

Finally, we obtain a U(1) gauge invariant shift vector analogous to the geodesic curvature $\kappa_g$

$$R_{mn}^{a,b}(\bm{k}) = \partial_{k_a} \arg\langle\widetilde{n}|\partial_{k_b}|\widetilde{m}\rangle = \mathcal{A}_m^a(\bm{k}) - \mathcal{A}_n^a(\bm{k}) - \partial_{k_a} \arg r_{mn}^b(\bm{k}). \tag{5}$$

As shown in Fig 1(b), $|\widetilde{n}\rangle$ is parallel transported along $|\widetilde{m}\rangle$ in the Hilbert space. $\partial_k|\widetilde{m}\rangle$ serves as an analogy of the tangent vector. The shift vector corresponds to the $\bm{k}$-derivative of the geometrical angle $\theta(\bm{k}) = \arg\langle\widetilde{n}|\partial_k|\widetilde{m}\rangle$. It's evident that $R_{mn}^{a,b}(\bm{k}) = -R_{nm}^{a,b}(\bm{k})$ since $\arg\langle m|\partial_k|n\rangle = -\arg\langle n|\partial_k|m\rangle$. Similarly, the phonon drag shift vector is given by $R_{mn,v}^a(\bm{k},\bm{k}') = D_{\bm{k},\bm{k}'} \arg\langle\widetilde{n},\bm{k}|g|\widetilde{m},\bm{k}'\rangle$. This highlights the similarity between the shift vector and geodesic curvature.

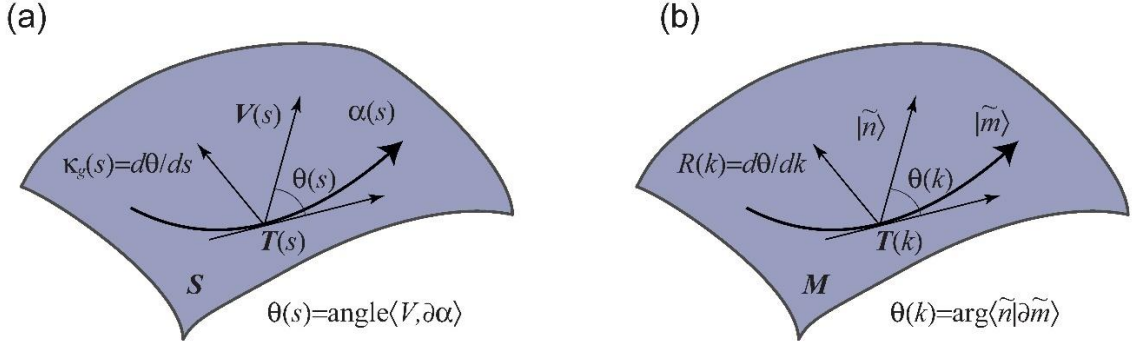

Figure 1. (a) Geodesic curvature of a curve α(s) on a 2D manifold $\bm{S}$ as a rate of turning. $\bm{V}(s)$ is the vector function lives in the tangent space, and is parallel transported along α(s). $\bm{T}(s)$ is the tangent vector and θ(s) is the angle between the vector function $\bm{V}(s)$ and tangent vector $\bm{T}(s)$. (b) Shift vector as a rate of turning on a 2D manifold $\bm{M}$. $|\widetilde{n}\rangle$ is parallel transported along $|\widetilde{m}\rangle$ in the Hilbert space. $\partial_k|\widetilde{m}\rangle$ represents the tangent vector, and $\theta(\bm{k}) = \arg\langle\widetilde{n}|\partial_k|\widetilde{m}\rangle$ is the argument of the inner product between the parallel transported vector and the tangent vector.

We employ a nondegenerate two-band model to elucidate the quantitative relationship between geodesic curvature, which can also be generalized to degenerate systems[14]. The generic Hamiltonian of such a system takes the form

$$H(\bm{k}) = d_x\sigma_x + d_y\sigma_y + d_z\sigma_z = \bm{d}(\bm{k})\cdot\bm{\sigma}, \tag{6}$$

where $\bm{\sigma}$ is the Pauli matrices. Due to the anticommutation relations of the Pauli matrices, the eigenvalues of the Hamiltonian are given by $\pm|\bm{d}(\bm{k})| = \pm\sqrt{d_x^2 + d_y^2 + d_z^2}$. A practical graphical representation of the Hamiltonian is the Bloch sphere, where the polar angle $\theta$ and azimuthal angle $\phi$ are defined as

$$\cos\theta = \frac{d_z}{|\mathbf{d}|}; e^{i\phi} = \frac{d_x + id_y}{|\mathbf{d}|}. \tag{7}$$

Starting with the rewritten Hamiltonian

$$H(\mathbf{k}) = |\mathbf{d}| \begin{pmatrix} \cos\theta & \sin\theta\, e^{-i\phi} \\ \sin\theta\, e^{i\phi} & -\cos\theta \end{pmatrix}, \tag{8}$$

we obtain the intra- and inter- Berry connection $\mathcal{A}_1^a = \partial_{k_a}\phi \sin^2\frac{\theta}{2}$, $\mathcal{A}_2^a = \partial_{k_a}\phi \cos^2\frac{\theta}{2}$, $A_{21}^a = \frac{i}{2}(\partial_{k_a}\theta - i\partial_{k_a}\phi \sin\theta)$. Subsequently, we have the shift vector

$$R_{21}^{a,b} = \mathrm{Im}\frac{A_{21}^b[\partial_{k_a}A_{12}^b - i(\mathcal{A}_1^a - \mathcal{A}_2^a)A_{12}^b]}{|A_{12}^b|^2}. \tag{9}$$

For simplicity, we consider $a = b$ and drop the $\mathbf{k}$ index in the partial derivative, yielding

$$R_{21} = \frac{\partial\theta\partial^2\phi \sin\theta + 2(\partial\theta)^2\partial\phi \cos\theta + (\partial\theta)^3 \sin^2\theta \cos\theta - \partial^2\theta\partial\phi \sin\theta}{(\partial\theta)^2 + (\partial\phi \sin\theta)^2}. \tag{10}$$

The geodesic curvature of the spherical curve $\mathbf{r}(\mathbf{k}) = \mathbf{d}(\mathbf{k})$ is given by[15]

$$\kappa_g = \left(\mathbf{r} \times \frac{d\mathbf{r}}{ds}\right) \cdot \frac{d^2\mathbf{r}}{ds^2} = \frac{\partial\theta\partial^2\phi \sin\theta + 2(\partial\theta)^2\partial\phi \cos\theta + (\partial\theta)^3 \sin^2\theta \cos\theta - \partial^2\theta\partial\phi \sin\theta}{[(\partial\theta)^2 + (\partial\phi \sin\theta)^2]^{\frac{3}{2}}}. \tag{11}$$

Additionally, we have the quantum metric which measures the distance between two adjacent Bloch states

$$g_{12} = A_{12}A_{21} = \frac{(\partial\theta)^2 + (\partial\phi \sin\theta)^2}{4}. \tag{12}$$

Finally, we establish the quantitative relationship among the geodesic curvature, shift vector, and quantum metric

$$\kappa_g = \frac{R_{21}}{2\sqrt{g_{12}}}. \tag{13}$$

The shift vector indeed serves as the geodesic curvature, quantifying the displacement by which the Bloch state of band $m$ deviates from that of band $n$ during an electric dipole transition within the $\mathbf{k}$-space manifold.

**Quantization and Wilson loop representation.** In a 2D compact Riemannian manifold $S$ with a smooth boundary $\partial S$, the Gauss-Bonnet theorem reads[26]

$$2\pi\chi(S) = \int_S G\, dA + \oint_{\partial S} \kappa_g\, ds, \tag{14}$$

where $G$ is the Gaussian curvature and $\chi$ denotes the Euler characteristic number. In quantum systems, during an adiabatic evolution along a closed $\boldsymbol{k}$-space loop $\partial M$ on the manifold within the Brillouin zone (BZ), a local gauge invariant quantized character is defined as

$$2\pi\chi_{mn} = \int_{M\subseteq BZ} \Omega_{mn}^z \, d^2k - \oint_{\partial M} \boldsymbol{R}_{mn} \cdot d\boldsymbol{k}, \tag{15}$$

where $\Omega_{mn}^a = \Omega_m^a - \Omega_n^a$ is the difference of intraband Berry curvature 2-form. Here, the index denoting the direction of the position operator in the shift vector is absorbed for brevity. The Berry curvature assumes the role of Gaussian curvature in the equation. This can be verified using Stokes' theorem, as evidenced by the equation $\Omega_{mn}^z = \partial_{\boldsymbol{k}} \times \boldsymbol{R}_{mn} \cdot \hat{\boldsymbol{z}}$[27]. The equivalence highlights the profound geometric connections between Berry curvature, shift vector, and traditional differential geometry concepts. We can generalize the above Gauss-Bonnet theorem in quantum system

$$2\pi\chi_{mn} = \int_{M\subseteq BZ} \partial_{\boldsymbol{k}} \times \boldsymbol{\mathcal{O}}_{mn} \cdot \hat{\boldsymbol{z}} \, d^2k - \oint_{\partial M} \boldsymbol{\mathcal{O}}_{mn} \cdot d\boldsymbol{k}, \tag{16}$$

where $\boldsymbol{\mathcal{O}}_{mn}$ is arbitrary intra-/inter- band connection or quantum geometric potential, which may be either gauge dependent or gauge invariant. If we choose the matrix element $\boldsymbol{\mathcal{O}}_{mn}$ to be gauge dependent Euler connection $\boldsymbol{\mathcal{O}}_{mn} = \boldsymbol{a}(\boldsymbol{k}) = \langle m|\partial_{\boldsymbol{k}}|n\rangle$, we recover the Euler class of a specific two-band subspace[13]. The concepts of Euler form and Euler class can be viewed as more nuanced counterparts of Berry curvature and Chern numbers, respectively. When $\boldsymbol{\mathcal{O}} = \boldsymbol{\mathcal{A}}_n$ is the intraband Berry connection, the inclusion of the additional Berry phase $\phi_n = \oint_{\partial M} \boldsymbol{\mathcal{A}}_n \cdot d\boldsymbol{k}$ along an adiabatic path encircling the Fermi surface leads to the nonquantized part of the intrinsic Hall conductivity[28], which is analogous to the $\boldsymbol{k}$-space manifestation of the Aharonov–Bohm effect. Next, we present the Wilson loop representation of the Gauss-Bonnet theorem. As illustrated in Fig. 2(a), the Berry curvature in a discretized Brillouin zone can be computed by Fukui-Hatsugai-Suzuki method [29]. It is represented by the overlap matrix elements between Bloch wavefunctions $|\mathcal{U}\rangle$ at neighboring $\boldsymbol{k}$-points and given by $\Omega_n^c(\boldsymbol{k}) = \arg W_n(\boldsymbol{k})$, where

$$\begin{aligned} W_n(\boldsymbol{k}) \equiv \epsilon_{abc} \langle n, \boldsymbol{k}|n, \boldsymbol{k}+\boldsymbol{q}_a\rangle \langle n, \boldsymbol{k}+\boldsymbol{q}_a|n, \boldsymbol{k}+\boldsymbol{q}_a+\boldsymbol{q}_b\rangle \\ \langle n, \boldsymbol{k}+\boldsymbol{q}_a+\boldsymbol{q}_b|n, \boldsymbol{k}+\boldsymbol{q}_b\rangle \langle n, \boldsymbol{k}+\boldsymbol{q}_b|n, \boldsymbol{k}\rangle, \end{aligned} \tag{17}$$

and $\boldsymbol{q}_a$ is an infinitesimal displacement vector along $a$ direction. Here, $\epsilon_{abc}$ denotes the Levi-Civita symbol. The Wilson loop formula of shift vector, as shown in Fig. 2(b), is expressed by[18]

$$R_{mn}^{a,b}(\boldsymbol{k}) = -\lim_{q_a\to 0} \partial_{q_a} \arg[W_{mn}(\boldsymbol{k}, \boldsymbol{q}_a, r^b, r^b)], \tag{18}$$

where the interband Wilson loop $W_{mn}$ is defined as

$$\begin{aligned} W_{mn}(\boldsymbol{k}, \boldsymbol{q}_a, r^b, r^b) \equiv \langle m, \boldsymbol{k}|m, \boldsymbol{k}+\boldsymbol{q}_a\rangle \langle m, \boldsymbol{k}+\boldsymbol{q}_a|r^b|n, \boldsymbol{k}+\boldsymbol{q}_a\rangle \\ \langle n, \boldsymbol{k}+\boldsymbol{q}_a|n, \boldsymbol{k}\rangle \langle n, \boldsymbol{k}|r^b|m, \boldsymbol{k}\rangle. \end{aligned} \tag{19}$$

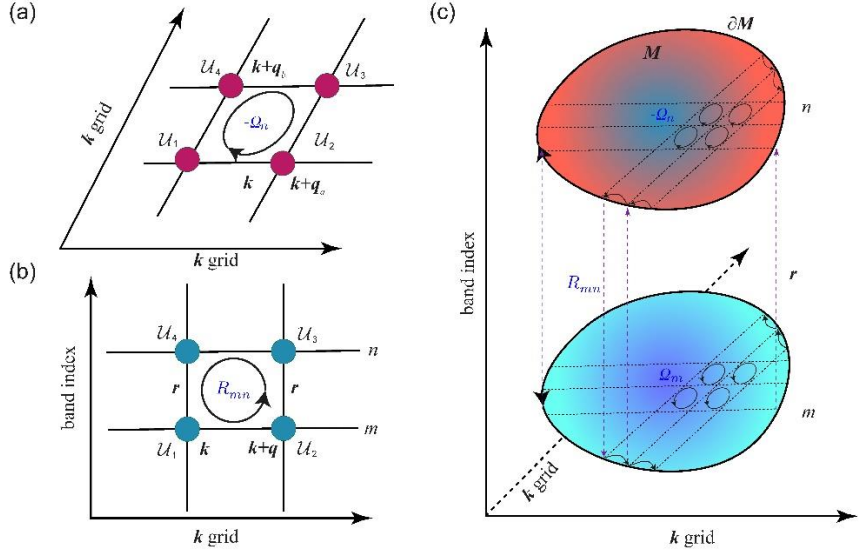

Figure 2. (a) Wilson representation of Berry curvature by the overlap matrix elements between Bloch wavefunctions $|\mathcal{U}\rangle$ at neighboring $\boldsymbol{k}$-points. (b) Wilson representation of shift vector by the overlap matrix elements and position matrix elements at neighboring $\boldsymbol{k}$-points. (c) Wilson loop representation of the quantized character $\chi_{mn}$, illustrating the summation of the intraband Berry curvature flux denoted by vector circles on surface $\boldsymbol{M}$ and the interband shift vector along the closed loop $\partial \boldsymbol{M}$.

The surface integral of Berry curvature on surface $\boldsymbol{M}$ and line integral of shift vector a closed $\boldsymbol{k}$-space loop $\partial \boldsymbol{M}$ can be reformulated to the sum of the Wilson loops. This transformation enables the Wilson representation of interband quantized character $\chi_{mn}$, as illustrated in Fig. 2(c). It's clear to see that $\chi_{mn}$ enumerates the singularities of the intraband Berry connection $\mathcal{A}_m^a(\boldsymbol{k})$ in the region $\boldsymbol{M}$ and the winding number $w = \oint_{\partial \boldsymbol{M}} \partial_{\boldsymbol{k}} \arg r_{mn}^b \cdot d\boldsymbol{k}$ of interband Berry connection $r_{mn}^b(\boldsymbol{k})$ along the closed loop $\partial \boldsymbol{M}$. Consequently, $\chi_{mn}$ remains quantized for all arbitrary $\boldsymbol{M}$ and their corresponding $\partial \boldsymbol{M}$. The shift vector can be extended to consider photon or phonon drag effects when perturbed by a wavevector deviating from the momentum of the photon or phonon, which connects two closed surfaces $\boldsymbol{M}$ and $\boldsymbol{M}'$, as depicted in Fig. 3(a). In the case of the phonon drag shift vector, we have

$$R_{mn,v}^a(\boldsymbol{k}, \boldsymbol{k}') = -\lim_{\boldsymbol{q}_a \to 0} \partial_{\boldsymbol{q}_a} \arg W_{mn,v}(\boldsymbol{q}_a, \boldsymbol{k}, \boldsymbol{k}'), \tag{20}$$

where the Wilson loop of the phonon drag shift vector, shown in Fig. 3(b), is given by

$$\begin{aligned} W_{mn,v}(\boldsymbol{q}_a, \boldsymbol{k}, \boldsymbol{k}') = &\langle m, \boldsymbol{k}|m, \boldsymbol{k}+\boldsymbol{q}_a\rangle\langle m, \boldsymbol{k}+\boldsymbol{q}_a|g|n, \boldsymbol{k}'\rangle \\ &\langle n, \boldsymbol{k}'|n, \boldsymbol{k}'-\boldsymbol{q}_a\rangle\langle n, \boldsymbol{k}'-\boldsymbol{q}_a|g|m, \boldsymbol{k}\rangle. \end{aligned} \tag{21}$$

Here, we obtain the quantized interband character of phonon drag shift vector

$$2\pi\chi_{mn} = \int_{M'}\Omega_m^z(\mathbf{k}')d^2k' - \int_M \Omega_n^z(\mathbf{k})d^2k - \oint_{\partial M,\partial M'} \mathbf{R}_{mn,v}(\mathbf{k},\mathbf{k}')\cdot d(\mathbf{k},\mathbf{k}'), \qquad (22)$$

where $d(\mathbf{k},\mathbf{k}') = d\mathbf{k} + d\mathbf{k}'$. Fig. 3(c) illustrates the quantization of phonon shift vector through Stokes' theorem. The phonon drag shift vector vanishes when $\mathbf{k} = \mathbf{k}'$.

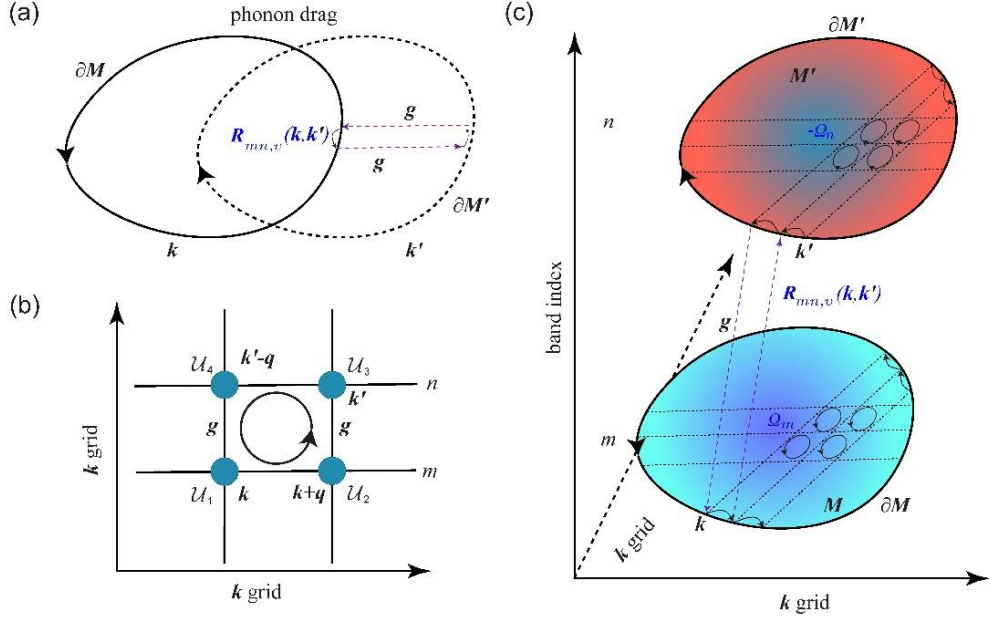

Figure 3. Phonon drag effect and quantization of boson drag shift vector. (a) Phonon drag shift vector. The two closed surfaces $\mathbf{M}$ and $\mathbf{M}'$ are connected by a phonon wavevector and electron-phonon coupling (b) Wilson representation of phonon drag shift vector by the overlap matrix elements and electron-phonon coupling matrix elements at neighboring $\mathbf{k}$-points. (c) Wilson loop representation of the quantized character $\chi_{mn}$ of phonon drag shift vector.

**2D Massive Dirac model and nonquantized part of CPGE.** To illustrate the concept of the character number and quantization with physical significance, we employ a simple two-dimensional (2D) massive Dirac model with two bands. This model describes massive Dirac fermions in 2D through the following Hamiltonian

$$d_x = k_x, d_y = k_y, d_z = m. \qquad (23)$$

The band energy surface and closed loop $\partial \mathbf{M}$ induced by circularly polarized light are shown in Fig. 4(a). Fig. 4(b) depicts the intraband Berry connection $\mathcal{A}_1$ and curvature $\Omega_1^z = \frac{1}{2}\epsilon_{\alpha\beta\gamma}d_\alpha\partial_{k_x}d_\beta\partial_{k_y}d_\gamma = \frac{m}{2d^3}$ of the conduction band. One can see that the curl of the gauge dependent intraband Berry connection directly yields the intraband Berry curvature. Furthermore, the shift vector can be determined by $R_{12}^{x,x} = 0$, $R_{12}^{y,x} = \frac{mk_x}{(k_x^2+k_y^2+m^2)^{1/2}(k_y^2+m^2)}$, $R_{12}^{x,y} = -\frac{mk_y}{(k_x^2+k_y^2+m^2)^{1/2}(k_x^2+m^2)}$, $R_{12}^{y,y} = 0$. The phase $\arg r_{12}^x$ of interband Berry connection $r_{12}^x$ is depicted in Fig. 4(c) and the corresponding shift vector field $\mathbf{R}_{12}^x = (R_{12}^{x,x}, R_{12}^{y,x})$ is illustrated in Fig. 4(d). The winding number $\oint_{\partial \mathbf{M}} d\arg r_{12}^x$ is determined by singularities of the phase along the loop $\partial \mathbf{M}$.

Similarly, the curl of the gauge invariant shift vector directly determines the interband Berry curvature. The surface integral of interband Berry curvature is $I_\Omega = \frac{1}{2\pi}\int_{|k|=k_c}\Omega^z_{12}\,d^2k = \text{sign}(m) - \frac{m}{(k_c^2+m^2)^{1/2}}$, and the loop integral of shift vector is given by $I_R = \frac{1}{2\pi}\oint_{\partial M} \boldsymbol{R}_{12}\cdot d\boldsymbol{k} = \frac{1}{2\pi}\int_0^{2\pi} \boldsymbol{R}_{12}(\boldsymbol{k}(\theta))\cdot\frac{d\boldsymbol{k}(\theta)}{d\theta}d\theta = -\frac{m}{(k_c^2+m^2)^{1/2}}$. Therefore, the interband character number, denoted by $\chi_{12} = \text{sign}(m)$, is twice the Chern number. Actually, in a general two-band model, the intraband Berry curvature is simply twice the interband Berry curvature[27]. Next, we discuss the physical meaning of the quantization. Under circularly polarized light with $\boldsymbol{E}(\omega)$+c.c., the CPGE injection current is denoted by $\frac{dj}{dt} = \eta_{ij}\boldsymbol{E}(\omega)\times\boldsymbol{E}^*(\omega)$. The trace of the conductivity tensor can be expressed as the Berry flux $\text{Tr}(\eta_{ij}) = \frac{ie^3}{2\hbar^2}\int_M d\boldsymbol{M}\cdot\boldsymbol{\Omega}$ in a general two-band model[10], where $d\boldsymbol{M}$ denotes the oriented surface element normal to $\boldsymbol{M}$. Consequently, if the integral is taken over a surface enclosing a Weyl node, the CPGE trace becomes exact quantized. In trivial systems, the circulation of shift vector contributes a non-quantized part to the trace, which serves as an interband equivalent to nonquantized part of the intrinsic Hall conductivity[28].

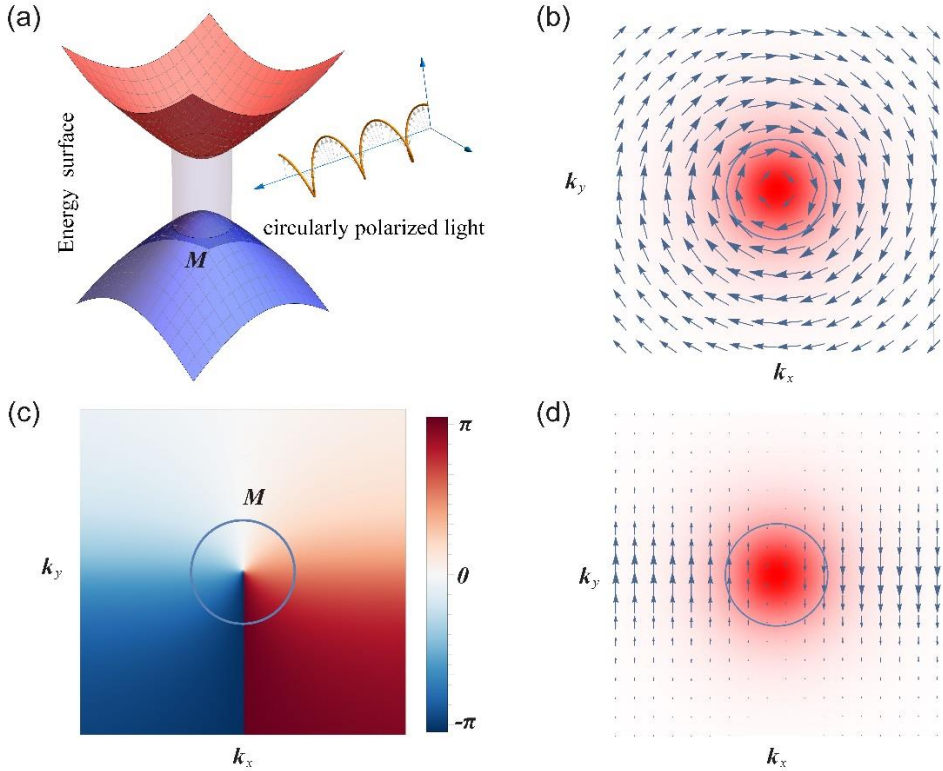

Figure 4. Two-band massive Dirac model. (a) Energy surfaces of conduction and valance bands. (b) Vector field of the intraband Berry connection $\mathcal{A}_1$ and intraband Berry curvature of the conduction band $\Omega^z_1$, indicated by arrows and red surface, respectively. (c) The phase $\arg r^x_{12}$ of interband Berry connection. (d) Shift vector $\boldsymbol{R}^x_{12}$ and interband Berry curvature $\Omega^z_{12}$, indicated by arrows and red surface, respectively.

**Conclusions**. In conclusion, we elucidated the geodesic nature and interband geometry of shift vector using the Wilson loop representation, which we verified in a two-band massive Dirac model. The quantization reveals a connection to the Gauss-Bonnet theorem applied to the intersubspace manifold. Furthermore, the shift vector acts like a quantum geometric potential, analogous to the intraband Berry connection. This analogy extends to the quantized character, which resembles the Chern number. It is noteworthy that the loop integral of the shift vector contributes to the non-quantized component of the trace of CPGE conductivity. The quantization of interband character provides valuable insights into the geometric foundation of these interband gauge invariant quantities and paves the way for a deeper understanding of their influence on nonlinear optical phenomena.

**Acknowledgements** We gratefully acknowledge helpful conversations with Congjun Wu and Jian Li from Westlake University, as well as Ji Feng from Peking University, Haiqing Lin from Zhejiang University, and Jian Zhou from Xi'an Jiaotong University. Hua Wang acknowledges the support from the National Natural Science Foundation of China (NSFC) under Grant No. 12304049. Kai Chang acknowledges the support from the Strategic Priority Research Program of the Chinese Academy of Sciences (Grants Nos. XDB28000000 and XDB0460000), the NSFC under Grant No. 92265203, and the Innovation Program for Quantum Science and Technology under Grant No. 2024ZD0300104.